\documentclass[twocolumn,showpacs,prb,aps,superscriptaddress]{revtex4}
\usepackage[latin9]{inputenc}
\usepackage{amsmath}
\usepackage{amssymb}
\usepackage{graphicx}
\usepackage{esint}

\makeatletter
\@ifundefined{textcolor}{}
{%
 \definecolor{BLACK}{gray}{0}
 \definecolor{WHITE}{gray}{1}
 \definecolor{RED}{rgb}{1,0,0}
 \definecolor{GREEN}{rgb}{0,1,0}
 \definecolor{BLUE}{rgb}{0,0,1}
 \definecolor{CYAN}{cmyk}{1,0,0,0}
 \definecolor{MAGENTA}{cmyk}{0,1,0,0}
 \definecolor{YELLOW}{cmyk}{0,0,1,0}
}

\makeatother

\begin{document}

\title{A magnetic monopole in topological insulator: exact solution}

\author{Yuan-Yuan Zhao and Shun-Qing Shen}

\affiliation{Department of Physics, The University of Hong Kong, Pokfulam Road,
Hong Kong, China}
\begin{abstract}
We present an exact solution of a magnetic monopole in a topological
insulator. It is found that a magnetic monopole can induce a pair
of zero energy modes: one is bound to the monopole and the other is
distributed near the surface. For a finite size system, the interference
of two states may lift the degeneracy, and the resulting states have
one half near the origin and another half around the surface. However,
the energy difference decays exponentially with the size of the system.
Due to the particle-hole symmetry only one electron can occupy the
two states at half filling. The presence of a pair of degenerate zero
energy modes does not fully support the realization of the Witten
effect in a topological insulator as the charge bound to the monopole
can be from zero to one modulus an integer. External fields such as
the Zeeman field may remove the degeneracy of two states. In this
case, a half charge around the monopole becomes possible.
\end{abstract}

\pacs{03.65.Vf, 14.80.Va, 14.80.Hv}

\maketitle
Topological insulators are electronic materials that behave like insulators
or semiconductors in the bulk, but are surrounded by a topologically
protected conducting layer near the surface of the materials \cite{Hasan-10rmp,Qi-11rmp,Moore}.
One of the predicted features in the materials is the ``axion electrodynamics''
\cite{Qi-98prb}, as a response to external electromagnetic fields.
The idea of axion was first introduced to address the strong charge-parity
problem in the physics of strong interaction \cite{Pecci-77prl}.
It becomes a possible candidate for the dark matter in the universe,
and however, has not been confirmed yet experimentally so far. Historically,
it was known that an additional term $\theta\frac{e^{2}}{2\pi h}\mathbf{B}\cdot\mathbf{E}$
can be introduced into the Maxwell Lagrangian for electric and magnetic
fields, which is time reversal invariant when $\theta=\pi$. This
additional term revises both the Gauss' law and Ampere's law in the
Maxwell's equations by adding extra terms \cite{Wilczek-87prl},
\begin{equation}
\nabla\cdot\mathbf{D=\rho_{e}-\frac{\alpha}{\pi\mu_{0}c}}\nabla\theta\cdot\mathbf{B},
\end{equation}
\begin{equation}
\nabla\times\mathbf{H}=\partial_{t}\mathbf{D}+j+\frac{\alpha}{\pi\mu_{0}c}\left(\nabla\theta\times\mathbf{E}+\partial_{t}\theta\mathbf{B}\right)
\end{equation}
where $\mathbf{D}=\epsilon_{0}\mathbf{E}+\mathbf{P}$, $\mathbf{H}=\frac{1}{\mu_{0}}\mathbf{B}-\mathbf{M}$,
and $\alpha=\frac{e^{2}}{2\epsilon_{0}hc}$ is the fine structure
constant. One of the fundamental properties in the revised Maxwell's
or axion equations is the Witten effect, which states that a magnetic
monopole of unit strength $e{}_{M}=\phi_{0}=\frac{h}{e}$ in an axion
media must bind an electric charge, $-(n+\frac{\theta}{2\pi})e$,
where $e(>0)$ is the elementary charge and $n$ is an integer. Consider
a point-like magnetic monopole situated at the origin, which produces
a magnetic field, $\nabla\cdot\mathbf{B}=\phi_{0}\delta(\mathbf{r})$.
We suppose that $\theta=0$ initially and then increases adiabatically
to $\theta=\pi$. $\theta$ is uniform in the space and there is no
current in the media. It follows from the axion equations that
\begin{equation}
\delta\rho_{e}=\rho_{e}(\theta=\pi)-\rho_{e}\left(\theta=0\right)=-\frac{\alpha}{\mu_{0}c}\nabla\cdot\mathbf{B}=-\frac{e}{2}\delta(\mathbf{r}).
\end{equation}
As a magnetic monopole does not induce a half elementary charge in
a conventional media of $\theta=0$, the charge bound to the monopole
should be $-e/2$ modulus an integer for a time reversal invariant
topological insulator of $\theta=\pi$ \cite{Witten-79plb}. Charge
fractionalization in condensed matters was extensively discussed for
one dimension in 1980s \cite{Jackiw-76prd,Kivelson-82prb,Heeger-88rmp}.
The quasi-particles in the fractional quantum Hall effect also carry
fractional charge \cite{Laughlin-82prl,Jain-88prl}. Whether or not
a half elementary charge bound by a magnetic monopole could exist
in a topological insulator becomes a subtle issue to test the validity
of the axion theory for topological insulators. Rosenberg and Franz
studied the Witten effect in a crystalline topological insulator numerically
\cite{Rosenberg-10prb}, and intended to use it as a criterion to
justify whether the system is topologically trivial or non-trivial
\cite{Guo-10prl}.

In this paper, we present an exact solution of a magnetic monopole
located in the center of a topological insulator sphere, which is
described by the modified Dirac-like equation. It is found that there
exist a pair of degenerate solutions of zero energy: one is located
in the vicinity of the magnetic monopole and the other around the
sphere surface, which is characteristic of non-trivial topological
insulator. At half filling, only one electron occupies the two degenerate
states due to the particle-hole symmetry in the system. The double
degeneracy of the zero energy states does not favor or disfavor the
Witten effect because the bound charge near the monopole can be from
zero to $-e$ modulus an integer charge $-ne$, although it does not
exclude a half elementary charge as the Witten effect requires. The
degeneracy of the two states can be removed due to the finite size
effect for a small sphere. In the case the states are split into two
halves, one half is in the vicinity of the monopole, and another half
is distributed around the sphere surface. External fields such as
the Zeeman splitting can also remove the degeneracy.

The model Hamiltonian for a magnetic monopole in the modified Dirac
equation is given by \cite{Shen-11spin}

\begin{equation}
H=\left(\begin{array}{cc}
mv^{2}-B\Pi^{2} & v\sigma\cdot\Pi\\
v\sigma\cdot\Pi & -mv^{2}+B\Pi^{2}
\end{array}\right)\label{Dirac}
\end{equation}
where $2mv^{2}$ is the energy gap between the conduction band and
valence band, $v$ is the effective velocity and $B$ is a parameter
of dimension of inverse mass. $\sigma_{x,y,z}$ are the Pauli matrices.
The canonical momentum operator $\Pi=-i\hbar\nabla+e\mathbf{A}$ and
$\nabla\times\mathbf{A}=\frac{2q}{4\pi}\phi_{0}\mathbf{r}/r^{3}$.
$q$ is half of an integer due to the quantization of magnetic charge
and $q=\frac{1}{2}$ for a unit monopole \cite{Dirac-31}. It is known
that the vector potential $\mathbf{A}$ cannot be written as a single
expression in the whole space, and has to be defined as two functions
in two overlapping regions to keep singular \cite{Sakurai-book}.
In the absence of the magnetic monopole, the topological properties
of this equation have been understood very well. It is topologically
non-trivial for $mB>0$ and trivial for $mB<0$. A topological quantum
phase transition occurs at $mB=0$ \cite{Shen-11spin,Lu-10prb}. In
the presence of the magnetic monopole, the orbital angular momentum
is modified to $\mathbf{L}=\mathbf{r}\times\Pi-q\hbar\frac{\mathbf{r}}{r}$,
which satisfies the algebra $[\mathbf{L}_{\alpha},\mathbf{L}_{\beta}]=i\hbar\epsilon_{\alpha\beta\gamma}\mathbf{L}_{\gamma}$
\cite{Kazawa-77prd,book-monopole}. The eigenfunctions of $\mathbf{L}^{2}$
and $\mathbf{L}_{z}$ are denoted by $Y_{q,l,l_{z}}$ with the eigenvalues
$l(l+1)\hbar^{2}$ and $l_{z}\hbar$. Since the two terms in $\mathbf{L}$
are orthogonal to each other, $\mathbf{L}^{2}=\mathbf{|r}\times\Pi|^{2}+q^{2}\hbar^{2}$
and $l(l+1)\geq q^{2}$ . The total angular momentum is defined as
$\mathbf{J}=\mathbf{L}+\frac{\hbar}{2}\sigma$. The total angular
momentum $\mathbf{J}^{2}$ and its z-component $\mathbf{J}_{z}$ commute
with the Hamiltonian, and are good quantum numbers. Thus we can diagonalize
simultaneously $H$, $\mathbf{J}^{2}$ and $\mathbf{J}_{z}$. According
to the definition, $\mathbf{J}$ is the sum of two angular momenta,
$\mathbf{L}$ and $\frac{\hbar}{2}\sigma$. Thus the eigenvalues of
$\mathbf{J}^{2}$ can be $j=l+\frac{1}{2}$ and $j=l-\frac{1}{2}$,
respectively, and the corresponding eigenstates are the linear combination
of $Y_{q,l,l_{z}}$ and the eigenvectors for $\sigma_{z}$. For a
minimal $l(=|q|$) and $j=|q|-\frac{1}{2}$, the eigenstates are \cite{Kazawa-77prd,book-monopole}
\begin{equation}
\eta_{j,j_{z}}=\left(\begin{array}{c}
-\sqrt{\frac{|q|-m+\frac{1}{2}}{2|q|+1}}Y_{q,|q|,j_{z}-\frac{1}{2}}\\
\sqrt{\frac{|q|+m+\frac{1}{2}}{2|q|+1}}Y_{q,|q|,j_{z}+\frac{1}{2}}
\end{array}\right).
\end{equation}

Due to the particle-hole symmetry in the model Hamiltonian, the eigenvalues
of $\pm E$ appear in pairs. For a half-filled system, we are interested
in the energy levels near the Fermi surface at $E=0$ if existing.
Thus we focus on the case of $l=|q|$ and $j=|q|-\frac{1}{2}$ while
other cases of $j$ and $l$ may have the eigenstates below or above
than the Fermi level. To solve the problem, using the relations \cite{Kazawa-77prd},
$\left(\sigma\cdot\mathbf{r}\right)\eta_{j,j_{z}}=r\mathrm{sgn}(q)\eta_{j,j_{z}}$
and $\sigma\cdot\Pi f(r)\eta_{j,j_{z}}=-i\mathrm{sgn}(q)(\partial_{r}+r^{-1})f(r)\eta_{j,j_{z}}$,
where $f(r)$ are arbitrary function of $r$, we construct a simultaneous
eigenstate for $H$, $\mathbf{J}^{2}$ and $\mathbf{J}_{z}$:
\begin{equation}
\psi_{j,m}=\left(\begin{array}{c}
f(r)\eta_{j,j_{z}}\\
g(r)\eta_{j,j_{z}}
\end{array}\right)\equiv\left(\begin{array}{c}
f(r)\\
g(r)
\end{array}\right)\otimes\eta_{j,j_{z}}.
\end{equation}
Consequently the equation for the radial part of the wave function
is reduced into
\begin{equation}
\left\{ \left[1+\mathrm{sgn}(mB)\left(\partial_{\rho}^{2}-\frac{|q|}{\rho^{2}}\right)\right]\sigma_{z}-i\zeta\partial_{\rho}\sigma_{x}-\lambda\right\} \left(\begin{array}{c}
\rho f\\
\rho g
\end{array}\right)=0\label{eigen-eq}
\end{equation}
Here $\rho=kr$ ($k^{2}=|mv^{2}/B\hbar^{2}|$), $\zeta=\mathrm{sgn}(qmv)/\sqrt{|mB|}$
and $\lambda=E/mv^{2}.$

We first consider a large radius limit of the sphere, $\rho=kR\gg1.$
In this case it is reduced to a one-dimensional modified Dirac equation
by ignoring the term of $\frac{q}{\rho^{2}}$ at the end of $r=R$,
which has a solution of zero energy near $r=R$ when $mB>0$ \cite{Shen-11spin}.
In general, assume that $mB>0$ and $\lambda=0$. One obtains a general
solution

\begin{equation}
\psi_{j,m}^{s}=\chi_{s}\otimes\eta_{j,j_{z}}f_{s}(\rho)
\end{equation}
with $\sigma_{y}\chi_{s}=s\chi_{s}$ where $\chi_{s}^{T}=\frac{1}{\sqrt{2}}\left(1,is\right)$
and ( $s=\pm1$). For $\zeta^{2}\neq4$
\begin{equation}
f_{s}(\rho)=C_{1}e^{-s\zeta\rho/2}\frac{1}{\sqrt{\rho}}J_{\alpha}\left(\beta\rho\right)+C_{2}e^{-s\zeta\rho/2}\frac{1}{\sqrt{\rho}}K_{\alpha}\left(\beta\rho\right).
\end{equation}
where $J_{\alpha}(x)$ and $K_{\alpha}(x)$ ($\alpha=\sqrt{|q|+\frac{1}{4}}$
and $\beta=\sqrt{1-\zeta^{2}/4}$) are the Bessel functions of the
first and second kind, and $C_{i}$ ($i=1,2,3,4$) are the normalization
constants. For $\zeta^{2}>4$, $\beta$ becomes purely imaginary and
the solution is still valid. We can also use the modified Bessel functions
to replace the Bessel functions. Consider the boundary condition at
$\rho=0$ and $\rho=kR=\rho_{R}$. We have two solutions of zero energy.
For $s=1$ by taking $\varsigma>0$ without loss of generality, one
has a solution which is convergent at $\rho=0$, but decays exponentially
with a larger $\rho$
\begin{equation}
f_{+}(\rho)=C_{1}e^{-|\zeta|\rho/2}\frac{1}{\sqrt{\rho}}J_{\alpha}\left(\beta\rho\right).
\end{equation}
Thus the wave function denoted by $\psi_{j,m}^{+}(\rho)=\chi_{+}\otimes\eta_{m}f_{+}(\rho)$
is mainly located in the vicinity of the magnetic monopole. On the
other hand, both $J_{\alpha}(\beta\rho)$ and $K_{\alpha}(\beta\rho)$
become divergent for a large $\rho,$ but convergent for a small $\rho$.
For $s=-1$ one has the other solution which vanishes at $\rho_{R}=kR$
\begin{equation}
f_{-}(\rho)=C_{1}\frac{\sqrt{\rho_{R}}e^{|\zeta|\rho/2}}{\sqrt{\rho}e^{|\zeta|\rho_{R}/2}}\left(\frac{J_{\alpha}\left(\beta\rho\right)}{J_{\alpha}\left(\beta\rho_{R}\right)}-\frac{K_{\alpha}\left(\beta\rho\right)}{K_{\alpha}\left(\beta\rho_{R}\right)}\right).
\end{equation}
This solution denoted by $\psi_{j,m}^{-}(\rho)=\chi_{-}\otimes\eta_{m}f_{-}(\rho)$
is distributed near the sphere surface at $r=R$ and decays exponentially
in $\rho_{R}-\rho$ or $R-r$. This is a surface state of zero energy,
and is one of the characteristics of topological insulators. For $\zeta^{2}=4$
and $\beta=0$, we also have two solutions of zero energy: one is
near the origin,
\begin{equation}
f_{+}(\rho)=C_{3}e^{-|\zeta|\rho/2}\rho^{\alpha-1/2}
\end{equation}
with $s=1$ and the other is around the surface,
\begin{equation}
f_{-}(\rho)=C_{4}\frac{e^{|\zeta\rho/2}}{e^{|\zeta|\rho_{R}/2}}\left(\left(\frac{\rho}{\rho_{R}}\right)^{\alpha-\frac{1}{2}}-\left(\frac{\rho_{R}}{\rho}\right)^{\alpha+\frac{1}{2}}\right)
\end{equation}
with $s=-1$.

\begin{figure}
\includegraphics[width=8.5cm]{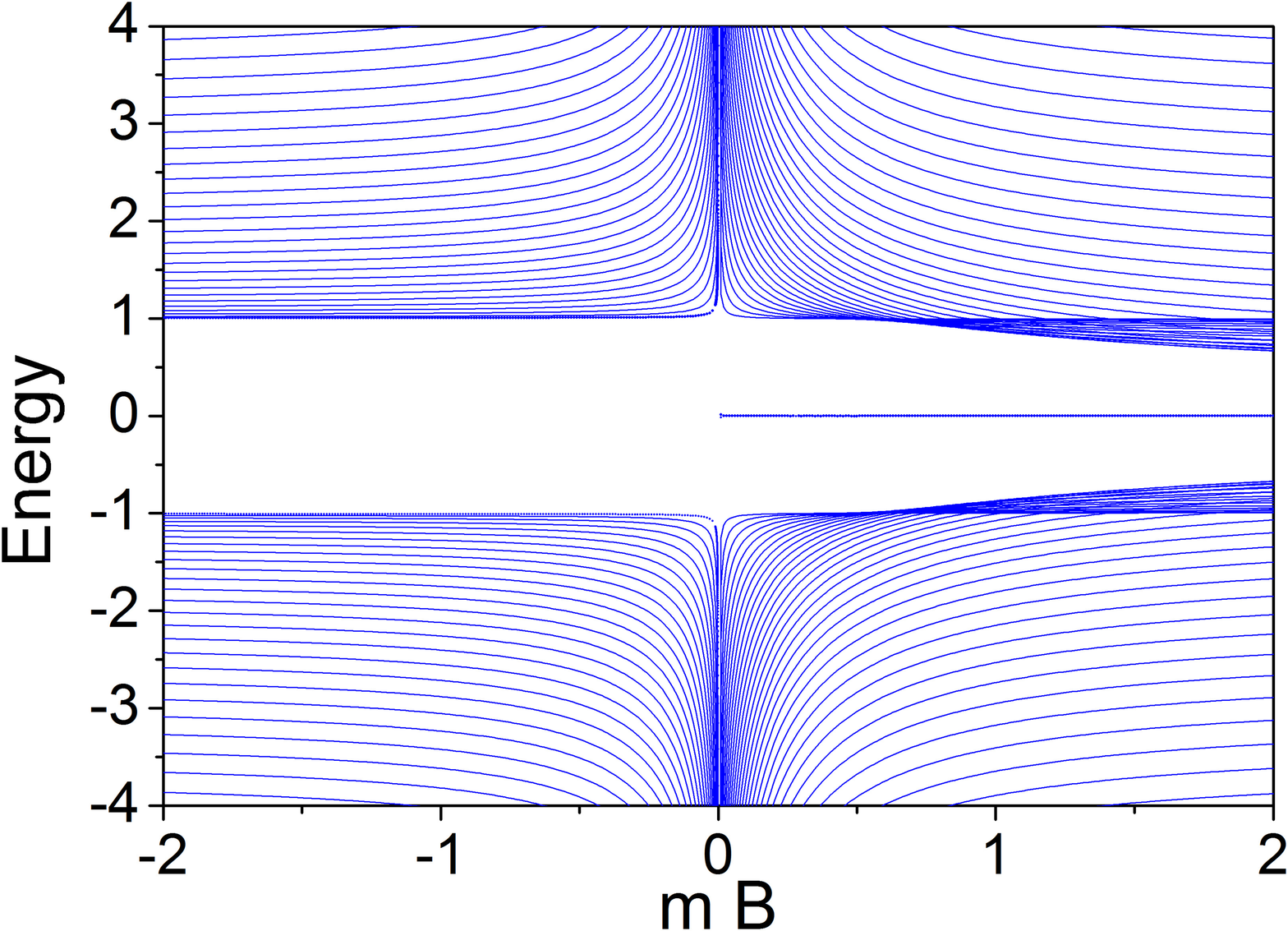}\caption{Energy spectrum of the bound states of $l=\left|q\right|=\frac{1}{2}$
and $j=l-\frac{1}{2}$ as a function of the dimensionless parameter
$mB$. The energy unit is $\left|m\right|v^{2}$. In general, the
states of the zero energy is $4\left|q\right|$-fold degeneracy.}
\end{figure}

\begin{figure}
\includegraphics[width=8.5cm]{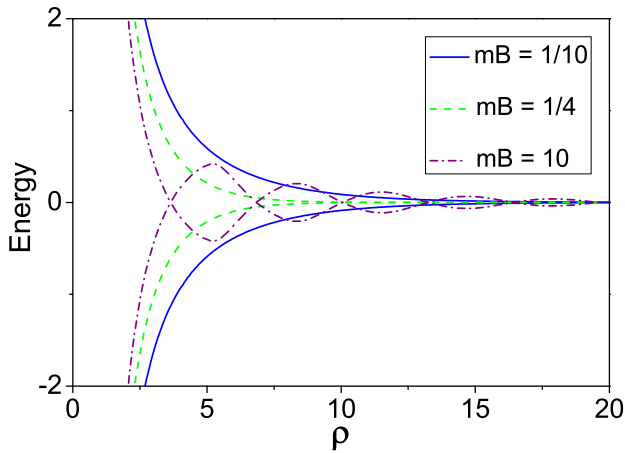}\caption{The energy splitting of the two bound states as a function of the
spherical radius $\rho_{R}=kR$. The energy unit is $\left|m\right|v^{2}$
and $k^{2}=\left|mv^{2}/B\hbar^{2}\right|$.}
\end{figure}

Except for the two solutions of zero energy, there also exist other
states even for $l=|q|$ and $j=|q|-\frac{1}{2}$. We solve Eq. (\ref{eigen-eq})
numerically, and present the results in Fig. 1. We note that the solutions
of zero energy exist only for the case of $mB>0$. There is no solution
of zero energy for $mB<0$ as the system is topologically trivial.
We find that all other states are not the surface states, some of
which are the bound states like those with different total quantum
numbers in a hydrogen atom. As the results are independent of $j_{z}$,
all these states are $2(2j+1)=4|q|$-fold degenerate. For $q=\pm\frac{1}{2}$,
the states are singlet with total angular momentum $j=0$. Due to
the topology of the band structures in topological insulator the existence
of the surface state should be robust against the continuous deformation
of the surface shape.

At half filling, the Fermi level is located at $E=0$ due to the particle-hole
symmetry in Eq.(\ref{Dirac}), below which all states or levels are
fully filled. Although we do not obtain general solutions of other
states of different $j$ and $j_{z}$, some of which are the surface
states between the energy gap, the doubly degenerate states at $E=0$
are closely related to possible realization of the Witten effect.
For $q=\frac{1}{2}$, only one electron occupies these two well-separated
and degenerate states. When $mB<0$, there is no solutions of zero
energy near the center and around the surface. Although it is possible
that the magnetic monopole can bind a lot of electron charges with
the energy below zero, the electron charges accumulated around it
must be an integer multiple of the elementary charge $-e$. This is
consistent with the fact that the system is topologically trivial
with $\theta=0$. When $mB>0$, the system is topologically non-trivial.
the appearance of the surface state of $E=0$ is one of the characteristics.
If the topological insulator is really an axion media, $\theta$ should
be $\pi$. Thus the sign change of $mB$ should accompany the change
of $\theta$ from 0 to $\pi$. Therefore if a topological insulator
is really an axion media, a half elementary charge must be bound to
the magnetic monopole as the Witten effect requires. However, the
double degeneracy of the zero energy solutions could make the charge
around the monopole from 0 to $-e$. Therefore our exact solutions
are neither in favor of nor against the picture of the Witten effect
as we couldn't exclude the possibility of a half elementary charge.

Now we come to consider a sphere of a finite radius $R$. Denote the
state near the origin by $\psi_{j,m}^{+}$ and the surface state by
$\psi_{j,m}^{-}$. If the radius is large enough such that the two
wavefunctions have no overlap in space, $\psi_{j,m}^{+}$ and $\psi_{j,m}^{-}$
are two exact solutions of zero energy. When the radius $R$ is finite
and the two wavefunctions overlap in space, the interference of the
two wavefunctions may lift the degeneracy of the two states, which
is dubbed the finite size effect \cite{Zhou-08prl}. As an degenerate
perturbation approach, we still use the two functions as the basis.
As $\chi_{s}$ are the eigenstates of $\sigma_{y}$, we have the relations
$\chi_{s}^{\dagger}\sigma_{z}\chi_{s}=\chi_{s}^{\dagger}\sigma_{x}\chi_{s}=0$.
Therefore the expectation values $\left\langle \psi_{j,m}^{\pm}|H|\psi_{j,m}^{\pm}\right\rangle \equiv0$.
However, $\chi_{s}^{\dagger}\sigma_{z}\chi_{-s}=2$ and $\chi_{s}^{\dagger}\sigma_{x}\chi_{-s}=-2is$,
then $\Delta=\left\langle \psi_{j,m}^{+}|H|\psi_{j,m}^{-}\right\rangle \neq0.$
As a consequence of the first-order degenerate perturbation, the energy
eigenvalues become $\pm|\Delta|$ and the two states become
\begin{equation}
\psi_{\pm}=\frac{1}{\sqrt{2}}\left(\psi_{j,m}^{+}\pm\frac{\Delta}{|\Delta|}\psi_{j,m}^{-}\right).
\end{equation}
The values of the $\pm|\bigtriangleup|$ are evaluated numerically
and plotted in Fig.2. There are two different cases. For $\varsigma^{2}\leq4$,
the gap increases monotonically with decreasing $R$ while for $\varsigma^{2}>4$
the gap oscillates, but the amplitude increases with decreasing $R$.
The states $\psi_{\pm}$ are separated into two halves as $\psi_{j,m}^{+}$
and $\psi_{j,m}^{-}$ are orthogonal and well separated in space.
The weights of these two parts are exactly equal to 1/2. When the
state with lower energy is occupied, the electron will be split into
two parts: one half is near the origin and the other is around the
surface as shown in Fig. 3.

To estimate the size of the wave package in real materials, we adopt
the fitted parameters from ARPES data for $\mathrm{Bi}{}_{2}\mathrm{Se}{}_{3}$
\cite{Zhang-10np}: $mv^{2}=0.126$eV, $B\hbar^{2}=21.8$eV$\mathrm{A}^{2}$,
and $\hbar v=2.94$eVA. As the wave function decays exponentially
combining a power law decay from the origin, $\mathrm{exp}[-r/\xi]$
or from the surface, $exp[-(R-r)/\xi]$, the characteristic length
is $\xi\simeq15$A. The finite size effect becomes obvious when $R$
is comparable with several $2\xi\simeq30$A, which is consistent with
the fact that gap opening in the $Bi_{2}Se_{3}$ thin film with thickness
several quintuple layers \cite{Zhang-10np}. Therefore the characteristic
length for a finite size effect is quite small.

Rosenberg and Franz claimed that a half electron charge is really
bound to a magnetic monopole in a crystalline topological insulator
\cite{Rosenberg-10prb}. By repeating their numerical calculation,
it is found that what they did is for the case of half filling minus
a single electron, i.e., the two zero energy modes are empty. It shows
that a half electron charge accumulating around the monopole originates
as the collective behaviors of the bulk states and the hydrogen-like
bound states around the magnetic monopole, not from a single particle
state splitting in the space. At the half filling, the charge accumulated
around the monopole are an integer for a finite size system, which
has no clue for the Witten effect. From our exact solutions, the double
degeneracy of zero energy solutions make the charge distribution around
the monopole uncertain: it can change from zero to one continuously.
For a small sphere, the degeneracy can be removed, and the the charge
accumulation near the monopole is an integer according to numerical
calculation. In this sense the exact solutions does not support the
Witten effect in topological insulator.

\begin{figure}[htbp]
\includegraphics[width=8.5cm]{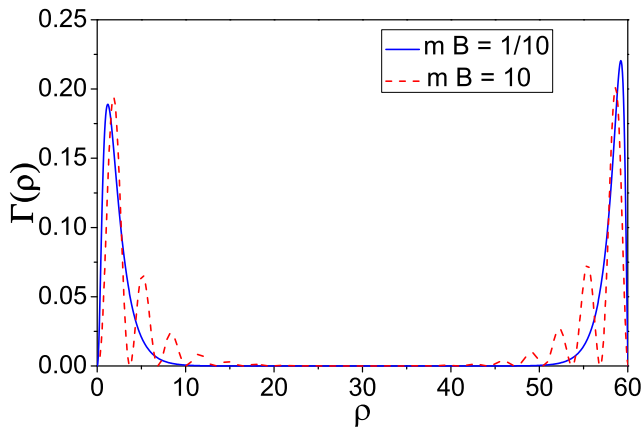} \caption{The radial probability density $\Gamma(\rho)=\rho^{2}\left(\left|f\right|^{2}+\left|g\right|^{2}\right)$
with a radius $\rho_{R}=kR=60$. The energy unit is $\left|m\right|v^{2}$
and $k^{2}=\left|mv^{2}/B\hbar^{2}\right|$.}
\end{figure}

Since a half electron charge is possibly accumulated near the monopole
in the case of half filling minus one electron, it is possible to
remove the degeneracy of the two zero energy modes by means of external
fields. Consider a radial magnetic field $\mathbf{B}=B_{r}\frac{\mathbf{r}}{r}$.
The Zeeman energy term is
\begin{equation}
\delta H=\mu_{eff}B_{r}\left(\begin{array}{cc}
\sigma_{r} & 0\\
0 & \sigma_{r}
\end{array}\right),
\end{equation}
which breaks the time reversal symmetry. $\sigma_{r}=\sigma\cdot\mathbf{r}/r$
is the Pauli operator along the radial direction. The energy shifts
of the two states of zero energy are $\Delta E_{\pm}=2\mathrm{sgn}(q)\mu_{eff}\int_{0}^{\rho_{R}}d\rho B_{r}|f_{\pm}(\rho)|^{2}$.
The degeneracy will be removed as the two integrals are usually not
equal. For example, if the field appears only near the surface due
to the proximity effect of a ferromagnetic layer, it is obvious that
$\Delta E_{+}=0$ and $\Delta E_{-}\neq0$. However, as the off-diagonal
integral is zero, $\left\langle \psi_{j,m}^{+}\right|\delta H\left|\psi_{j,m}^{-}\right\rangle =0$,
the two states will never be mixed as those due to the finite size
effect. Essentially the two states are still located near the origin
and the surface separately even in the presence of the Zeeman field.
Thus it becomes possible that the charge fractionalization may be
realized by the Zeeman field in the topological insulator or due to
the proximity effect of a ferromagnetic layer covering around the
surface.

In short we have found two exact solutions of zero energy modes for
a magnetic monopole with strength $\phi_{0}$ in a topological insulator:
one is located around the monopole and the other is distributed around
the surface. At half filling, only one electron occupies the two degenerate
states due to the particle-hole symmetry. Thus, the charge accumulated around the monopole is not determined
certainly, which is not in agreement with the prediction by the Witten
effect in an axion media. However, the external fields such as the
Zeeman field may remove the degeneracy of two zero energy modes. It
does not exclude the possibility that the half electron charge accumulates
around the monopole in a topological insulator.

We would like to thank Huai-Ming Guo for numerical calculations of
charge distribution in a lattice model. We also thank Shou-Cheng Zhang
for his critical comment. This work was supported by the Research
Grant Council of Hong Kong under Grant No.: HKU 7051/11P.

\end{document}